\documentclass{PoS}

\title{Quarkonium correlation functions at finite temperature in the charm to bottom region}

\ShortTitle{Quarkonium correlation functions at finite temperature in the charm to bottom region}

\author{\speaker{H.~Ohno} \\
	Physics Department, Brookhaven National Laboratory, \\
	Upton, NY 11973, USA \\
        E-mail: \email{hono@quark.phy.bnl.gov}}

\abstract{Quarkonium correlation functions at finite temperature were studied
	in a region of the quark mass for charmonia to bottomonia in quenched
	lattice QCD with $O(a)$-improved Wilson quarks. Our simulations were
	performed on large isotropic lattices at temperatures in the range
	from about 0.80$T_c$ to 1.61$T_c$. We investigated quarkonium behavior
        in terms of temperature dependence as well as quark mass dependence of
        the quarkonium correlation functions and related quantities at both vanishing
        and finite momenta.}

\FullConference{31st International Symposium on Lattice Field Theory LATTICE 2013 \\
		July 29 -- August 3, 2013 \\
		Mainz, Germany}

\begin{document}

\section{Introduction}
Quarkonium is an important probe to investigate properties of strongly interacting matters in hot medium.
In heavy ion collisions suppression of quarkonium production can signal formation of the deconfined quark and gluon phase called
quark-gluon plasma due to the color screening of the quark-antiquark potential \cite{Matsui:1986dk}.
In fact, suppression of yields of $J/\psi$ has been observed in nucleus-nucleus collisions at SPS \cite{Arnaldi:2009ph}, RHIC \cite{Adare:2008qa}
and LHC \cite{Abelev:2012rv,Aad:2010aa,Chatrchyan:2012np}.
Moreover, recently, the CMS collaboration has also reported suppression of the excited $\Upsilon$ state at LHC \cite{Chatrchyan:2012lxa}.
Because bottomonia are much heavier than charmonia, suppression signals of the bottomonia should suffer from smaller effects of cold nuclear
matters and are expected to be cleaner comparing to the charmonium case. Therefore, theoretical understanding of bottomonium
behavior at finite temperature is significant.

To study strongly interacting systems theoretically, a first principle calculation with lattice quantum chromodynamics (QCD)
is one of reliable approaches.
There are indeed many studies on charmonium behavior at finite temperature, which suggested that the S-wave states survive
up to about 1.5$T_c$, where $T_c$ is the critical temperature.
Some studies also indicated dissociation of the P-wave states just above $T_c$ \cite{Jakovac:2006sf,Aarts:2007pk,Ohno:2011zc}.
A most recent study with large fine lattices, however, showed a possibility of melting of the S-wave states
below 1.46$T_c$ \cite{Ding:2012sp} in contrast. For bottomonia, due to large mass of the bottom quark, only studies with
an effective field theory called the non-relativistic QCD have been done, which suggested survival of the S-wave states up to 2$T_c$
and dessication of the P-wave states immediately above $T_c$ \cite{Aarts:2010ek,Aarts:2012ka}. A study without effective theory thus is desirable.

In this study we investigated quarkonium behavior at finite temperature in a region of the quark mass for charmonia to bottomonia with quenched
lattice QCD simulation. Basically, the spectral function of the quarkonium has all the information of its in-medium properties and the spectral
function $\rho_H(\omega,\vec{p},T)$ at certain temperature $T$ is related to the Euclidean temporal correlator $G_H(\tau,\vec{p})$ as
\begin{equation}
G(\tau,\vec{p}) = \int^{\infty}_0 \frac{d\omega}{2\pi} \rho(\omega,\vec{p},T)\frac{\cosh[\omega(\tau-1/2T)]}{\sinh[\omega/2T]},
\end{equation}
where $G_H(\tau,\vec{p})$ is defined by
\begin{equation}
G(\tau,\vec{p}) \equiv \int d^3x \;e^{-i\vec{p}\cdot\vec{x}} \langle J(\tau,\vec{x})J^\dag(0,\vec{0}) \rangle.
\end{equation}
Here $J(\tau,\vec{x})\equiv \bar{\psi}(\tau,\vec{x})\Gamma\psi(\tau,\vec{x})$ is the meson operator and
$\Gamma=\gamma_5,\;\gamma_i,\;\textrm{\boldmath $1$},\gamma_5\gamma_i$ $(i=1,2,3)$ correspond to the pseudo-scalar (PS), vector (V), scalar (S) and
axial-vector (AV) channels, respectively. In general, to extract $\rho(\tau,\vec{p},T)$ from $G(\tau,\vec{p})$ is quite difficult since it requires many data points in the temporal
direction with statistically high precision. Therefore, as a first step, we focused on following two quantities instead of the spectral function itself:
One is the screening mass $M_{\mathrm{scr}}$ defined by the spatial correlator $G(z)$ as
\begin{equation}
G(z) \overrightarrow{_{z\to\infty}} \;e^{-M_{\mathrm{scr}}z},
\end{equation}
where $G(z)$ is also related to $\rho(\omega,\vec{p},T)$ as
\begin{equation}
G(z) = \int^{\infty}_0 \frac{2d\omega}{\omega}\int dp^3 \delta(p_x)\delta(p_y) \;e^{i\vec{p}\cdot\vec{x}}\rho(\omega,\vec{p},T).
\end{equation}
If there is a quarkonium ground state, the screening mass should be equal to corresponding mass
while, in the free quark case, it can be written by the lowest Matsubara frequency $\pi T$ and the quark mass $m_q$ as
$M_{\mathrm{scr}}=2\sqrt{(\pi T)^2+m^2_q}$. Another quantity investigated in this study is the reconstructed temporal correlator
$G_{\mathrm{rec}}(\tau,T;T')$, where momentum $\vec{p}$ is abbreviated. This quantity is defined by using the spectral function at temperature $T'$
and the integration kernel at $T$ as
\begin{equation}
G_{\mathrm{rec}}(\tau,T;T') \equiv \int^{\infty}_0 \frac{d\omega}{2\pi} \rho(\omega,T')\frac{\cosh[\omega(\tau-1/2T)]}{\sinh[\omega/2T]}.
\end{equation}
Especially in case that $T/T'=N'_\tau/N_\tau$ is some integer, where $N_\tau$ and $N'_\tau$ are temporal extents,
one can construct $G_{\mathrm{rec}}(\tau,T;T')$ with the conventional correlator $G(\tau,T')$ by
\begin{equation}
G_{\mathrm{rec}}(\tau,T;T') = \sum^{N'_\tau-N_\tau+\tau}_{\tau'=\tau;\Delta \tau'=N_\tau}G(\tau',T')
\end{equation}
\cite{Ding:2012sp}. Then the ratio $G(\tau,T)/G_{\mathrm{rec}}(\tau,T;T')$ reduces the influence of trivial $T$ dependence of the integration kernel and thus one can
discuss temperature dependence of the spectral function itself.

\section{Numerical results}

Our simulations were performed by using the standard plaquette gauge and $O(a)$-improved Wilson fermion actions with the quenched approximation.
The bare gauge coupling is $\beta=7.192$, which corresponds to the lattice spacing $a \simeq 0.0190$ fm ($a^{-1}\simeq 10.4$ GeV) determined from
Sommer scale 
$r_0=0.49$ fm. The lattice size is $96^3\times N_\tau$ with the temporal extent $N_\tau=48,\;32,\;28,\;24$ corresponding
to temperatures at 0.80$T_c$, 1.21$T_c$, 1.38$T_c$ and 1.61$T_c$, respectively, where $T_c\simeq 270$ MeV.
On each lattice the gauge configurations were generated with the pseudo-heatbath algorithm and after 2000 sweeps for the thermalization,
every 500 trajectories were stored for measurement. The temporal extent, corresponding
temperature and number of configurations are summarized in Table \ref{t1}. We chose 6 $\kappa$ values as listed in Table \ref{t2}, where corresponding
vector meson mass $m_V$ was given by the screening mass at $T=0.80T_c$. Here the screening mass was estimated by fitting the spatial vector correlator
to a single exponential. The largest and smallest $\kappa$ values were tuned to reproduce the experimental values of $J/\psi$ and $\Upsilon$ masses
\cite{Beringer:1900zz}, respectively.

\begin{table}[tbp]
\begin{center}
\caption{Temporal extent $N_\tau$, corresponding temperature $T$ in the unit of $T_c$ and number of configurations. \label{t1}}
\begin{tabular}{c|cccc}
\hline \hline
$N_\tau$  & 48   & 32   & 28   & 24   \\ \hline
$T/T_c$   & 0.80 & 1.21 & 1.38 & 1.61 \\
\# confs. & 259  & 476  & 336  & 336  \\
\hline \hline
\end{tabular}
\vspace{-1em}
\end{center}
\end{table}

\begin{table}[tbp]
\begin{center}
\caption{$\kappa$ value and corresponding vector meson mass $m_V$. \label{t2}}
\begin{tabular}{c|cccccc}
\hline \hline
$\kappa$    & 0.13194  & 0.13150  & 0.13100  & 0.13000  & 0.12800  & 0.12257  \\ \hline
$m_V$ [GeV] & 3.106(3) & 3.442(3) & 3.823(3) & 4.565(3) & 5.978(3) & 9.464(3) \\
\hline \hline
\end{tabular}
\vspace{-1em}
\end{center}
\end{table}

\begin{figure}[tbp]
 \begin{center}
  \includegraphics[width=49mm, angle=-90]{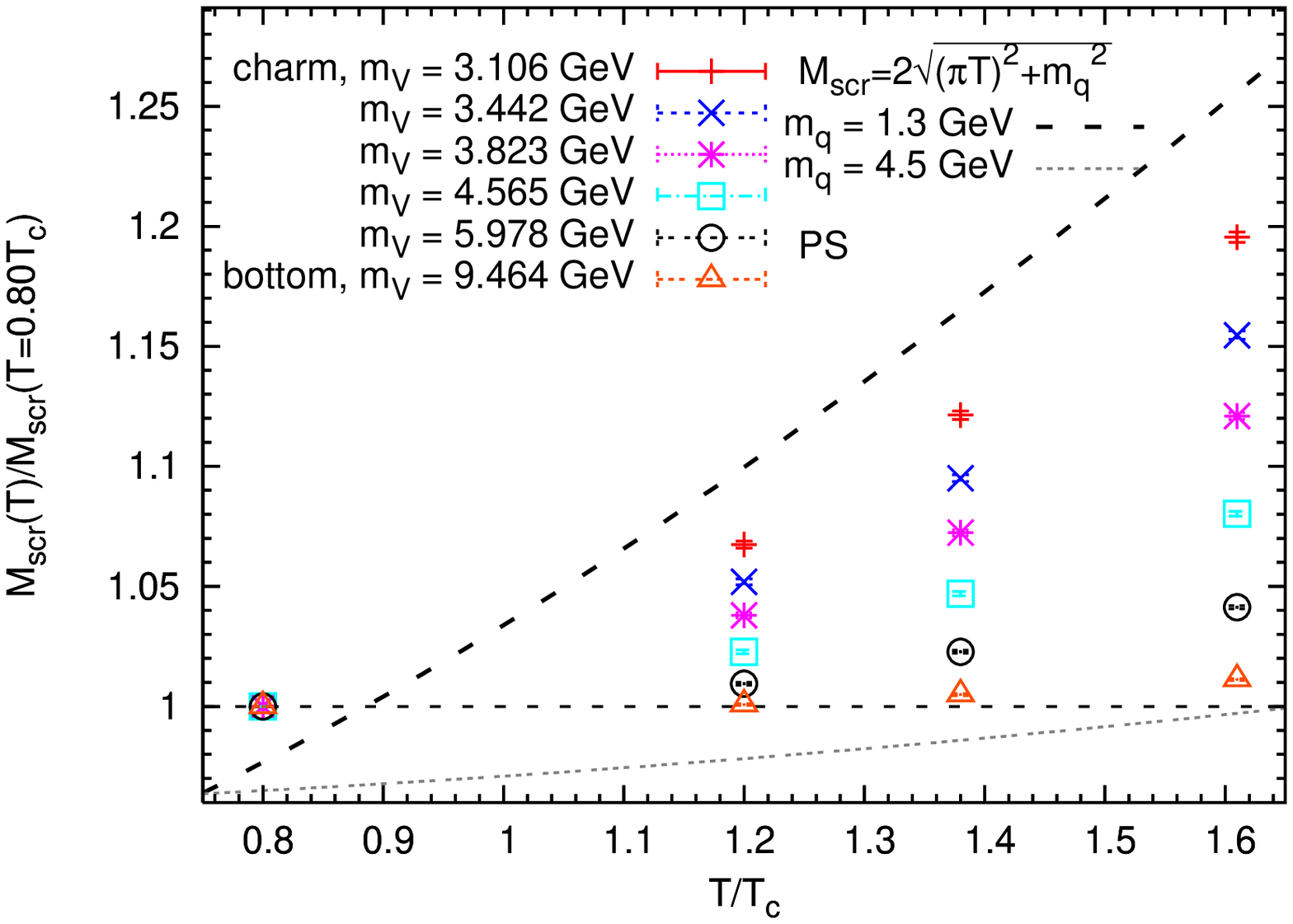}
  \includegraphics[width=49mm, angle=-90]{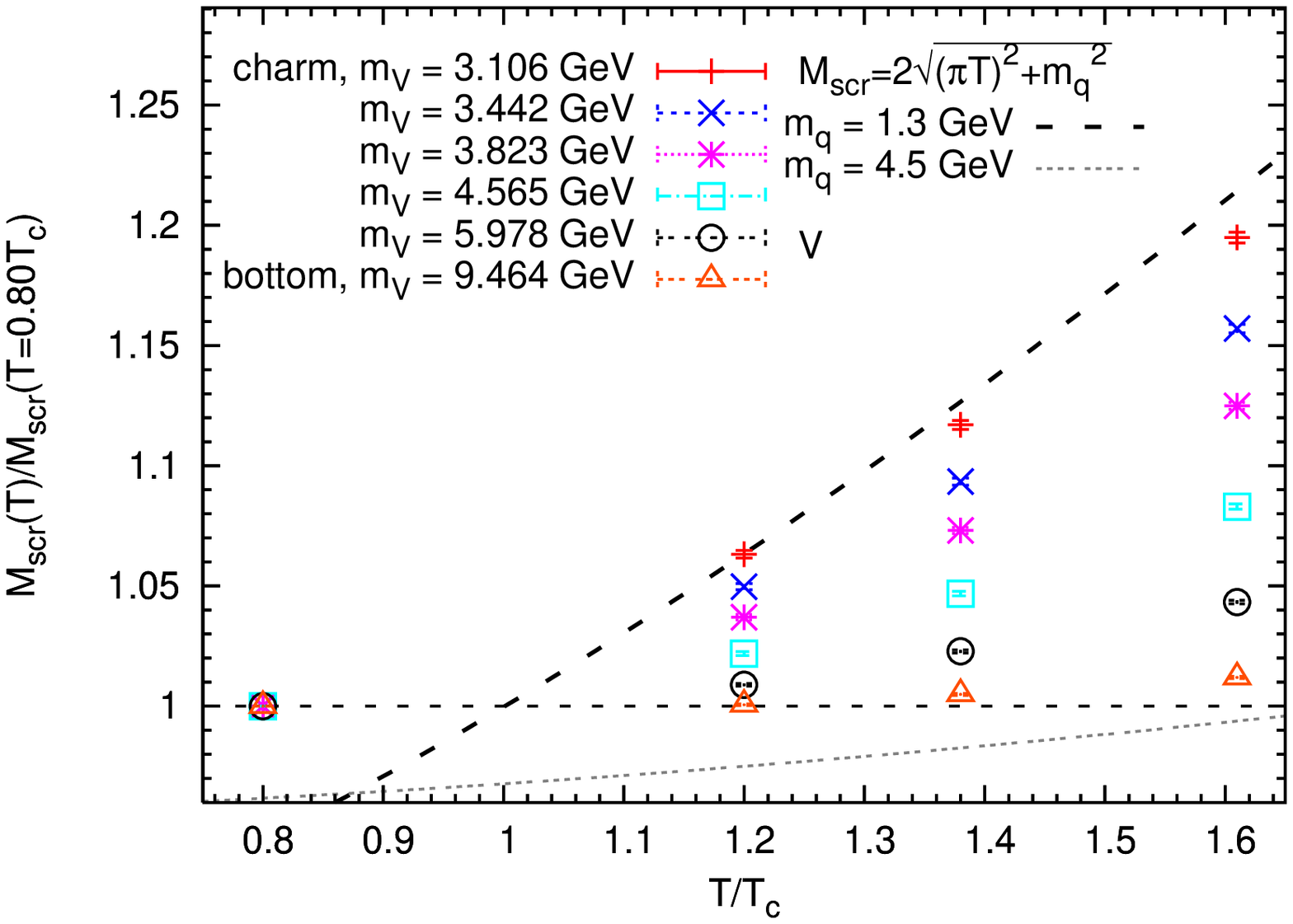}
  \includegraphics[width=49mm, angle=-90]{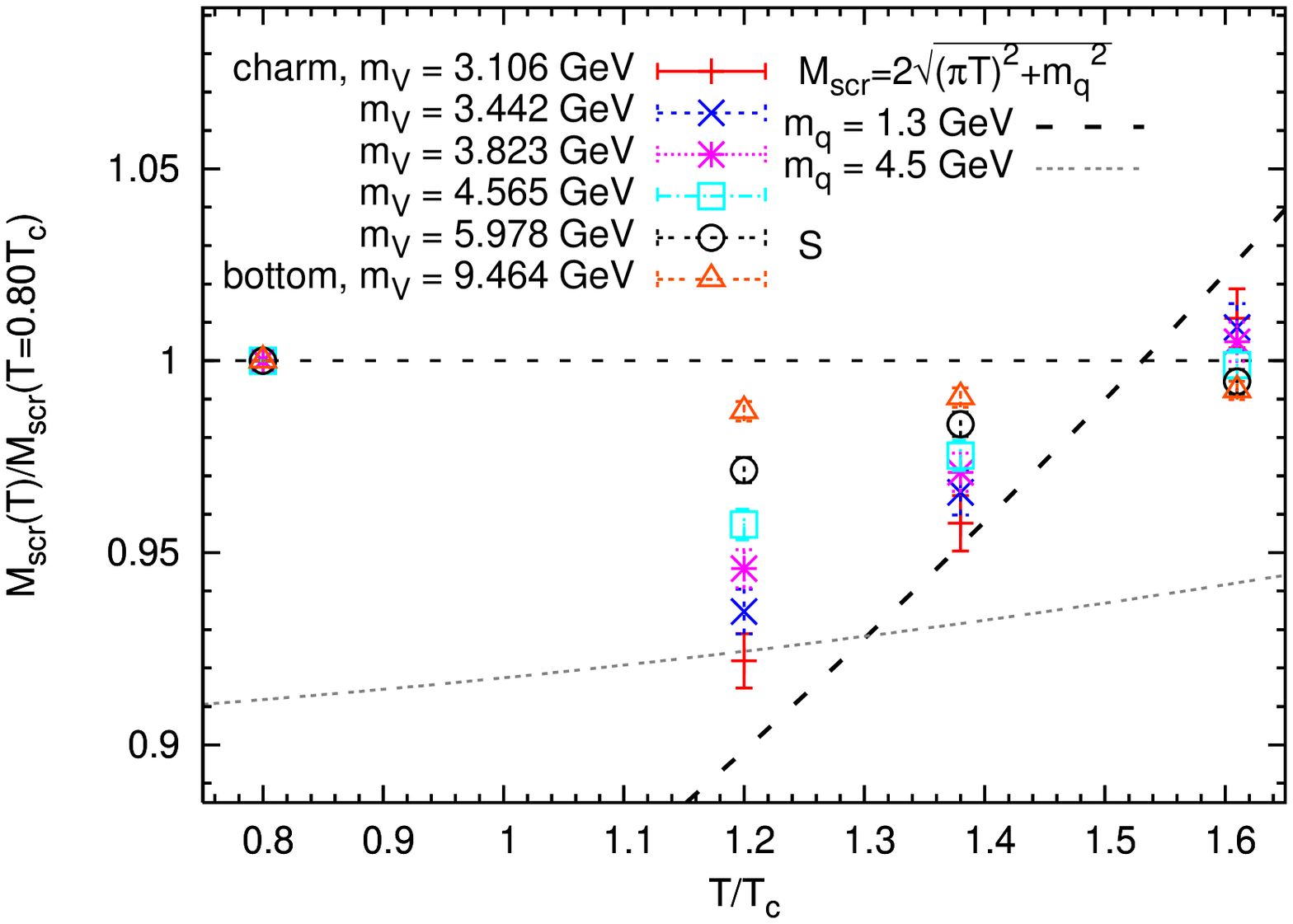}
  \includegraphics[width=49mm, angle=-90]{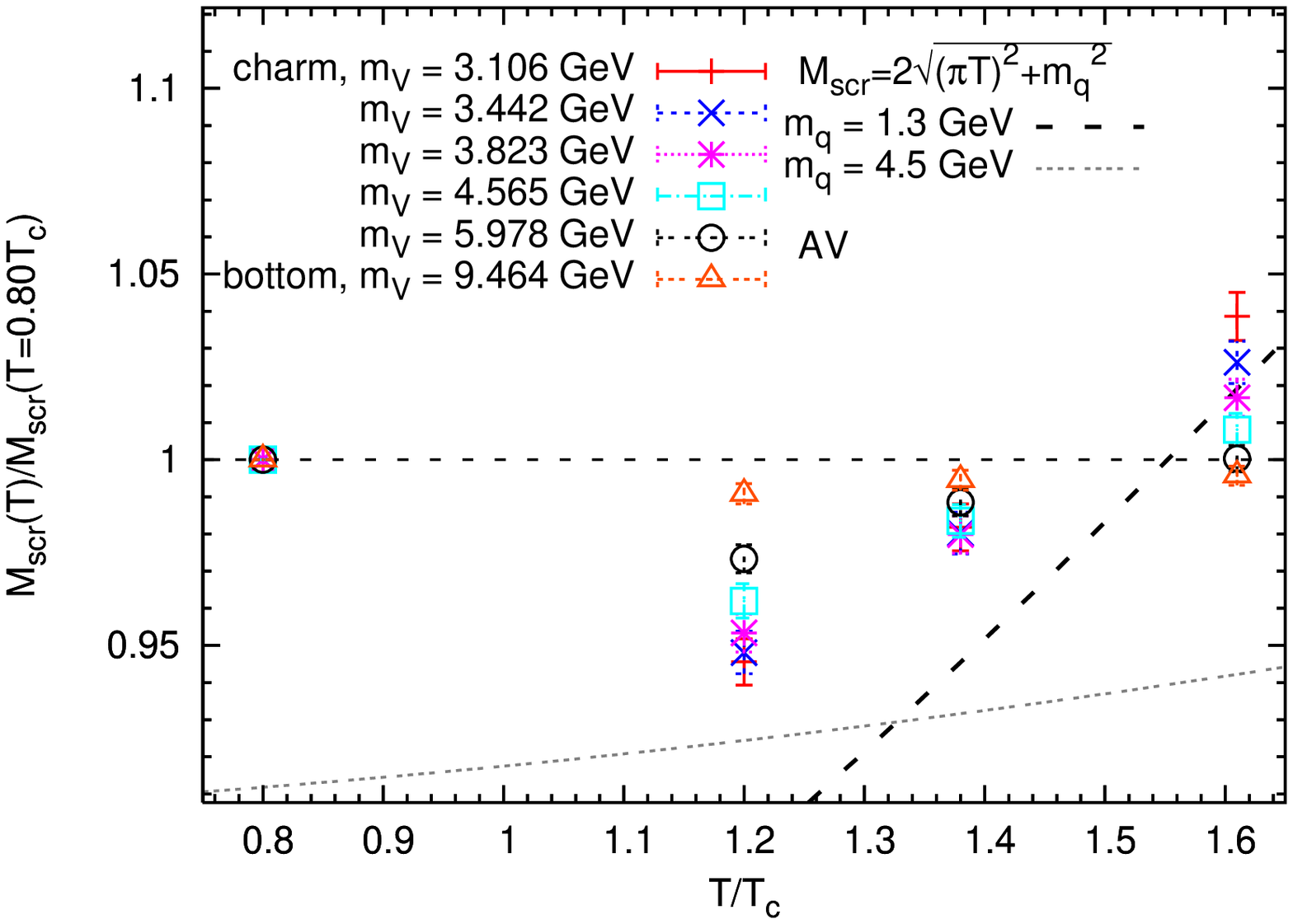}
  \caption{Temperature and quark mass dependence of the screening masses for PS (top-left), V (top-right), S (bottom-left) and AV (bottom-right)
  	channels. Each data is normalized by that for same quark mass and channel but at $T=0.80T_c$. Free quark cases with $m_q=1.3$ GeV
        and 4.5 GeV are also plotted with bold-dashed and dotted curves, respectively. \label{screening_mass}}
  \vspace{-1em}
 \end{center}
\end{figure}

In Figure \ref{screening_mass} temperature and quark mass dependence of the screening masses is shown, where each data is normalized by
that for same quark mass and channel but at $T=0.80T_c$. In case of the S-wave channels the screening mass increases as temperature increases
for all the quark masses and the temperature dependence is smaller for larger quark mass. Especially for the bottomonia the thermal effect
is negligibly small up to 1.21$T_c$ and only about 1\% even at 1.61$T_c$. This means that there is no clear evidence of the dissociation of
the S-wave bottomonium states at least up to 1.21$T_c$. On the other hand, the P-wave channels have quite different temperature dependence
for all quark masses comparing to the S-wave case, namely, the screening mass decreases above $T_c$ at first and then increases as increasing 
temperature. This suggests that not only the charmonia but also the bottomonia can have non-negligible thermal contribution just above $T_c$.
In addition, the non-monotonic temperature dependence seems to be related to the fact that the quarkonium mass can be larger than the screening
mass in the free quark case in the most of the temperature range investigated in this study, although one should carefully check the $m_q$
dependence.

\begin{figure}[tbp]
 \begin{center}
  \includegraphics[width=49mm, angle=-90]{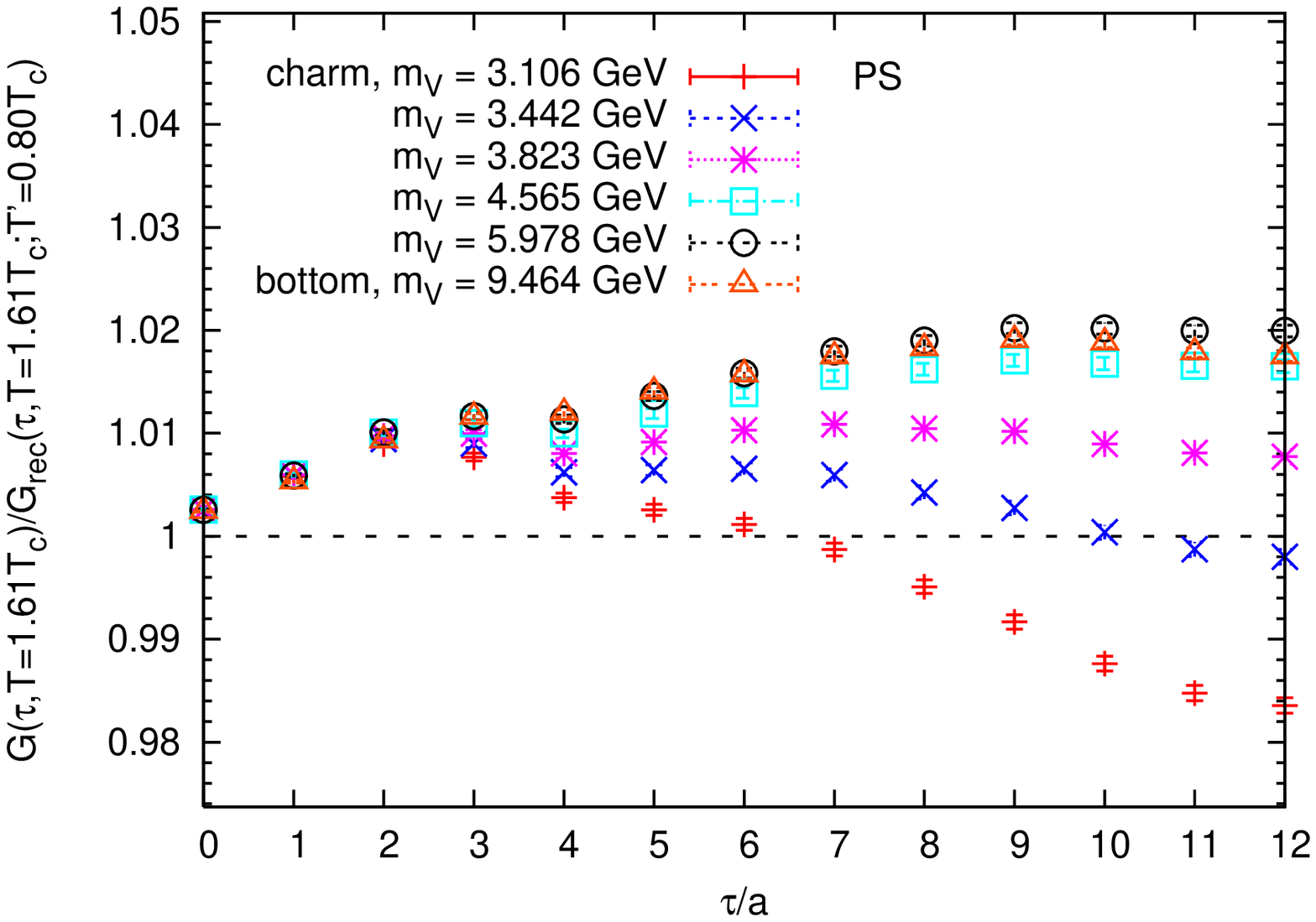}
  \includegraphics[width=49mm, angle=-90]{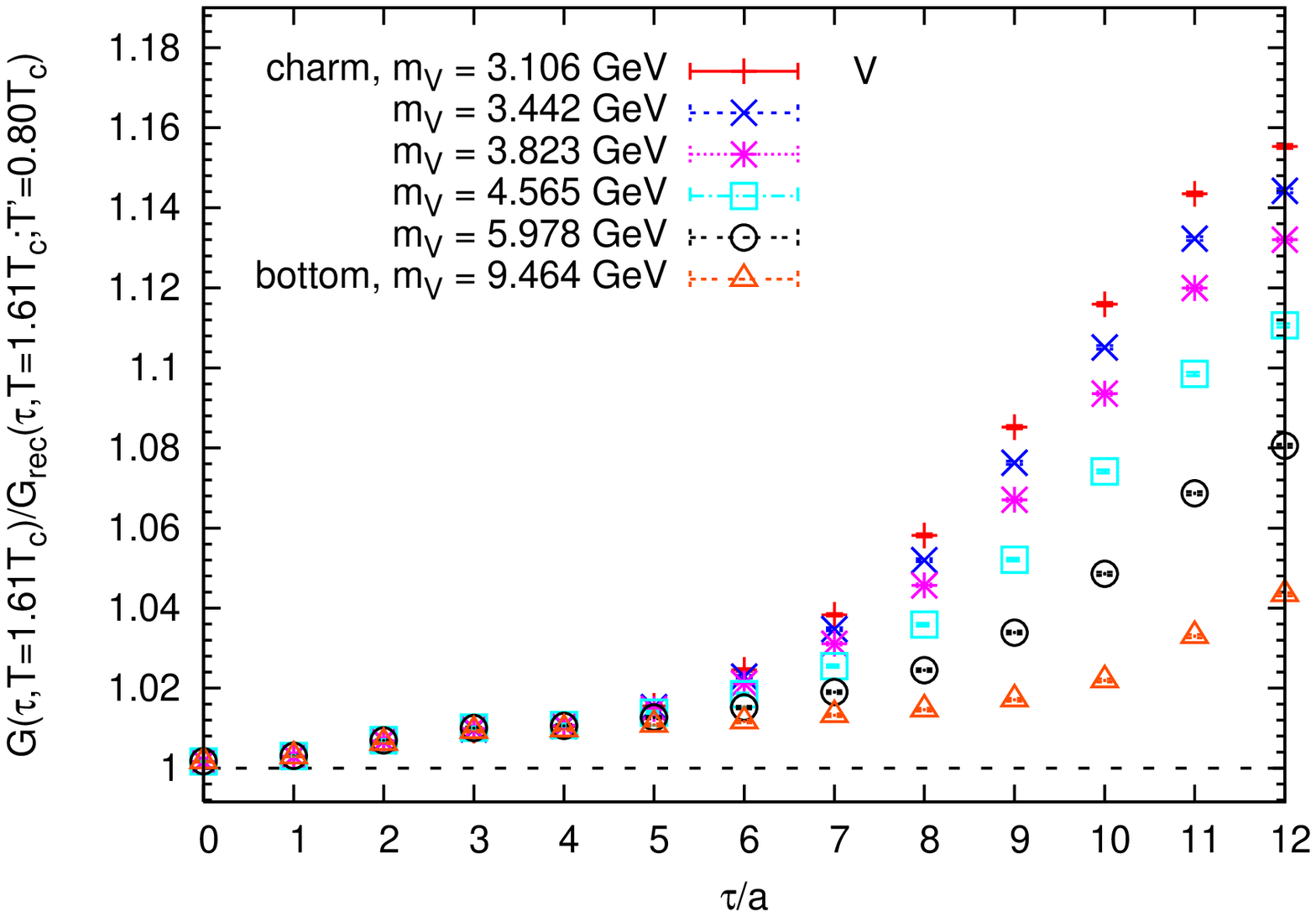}
  \includegraphics[width=49mm, angle=-90]{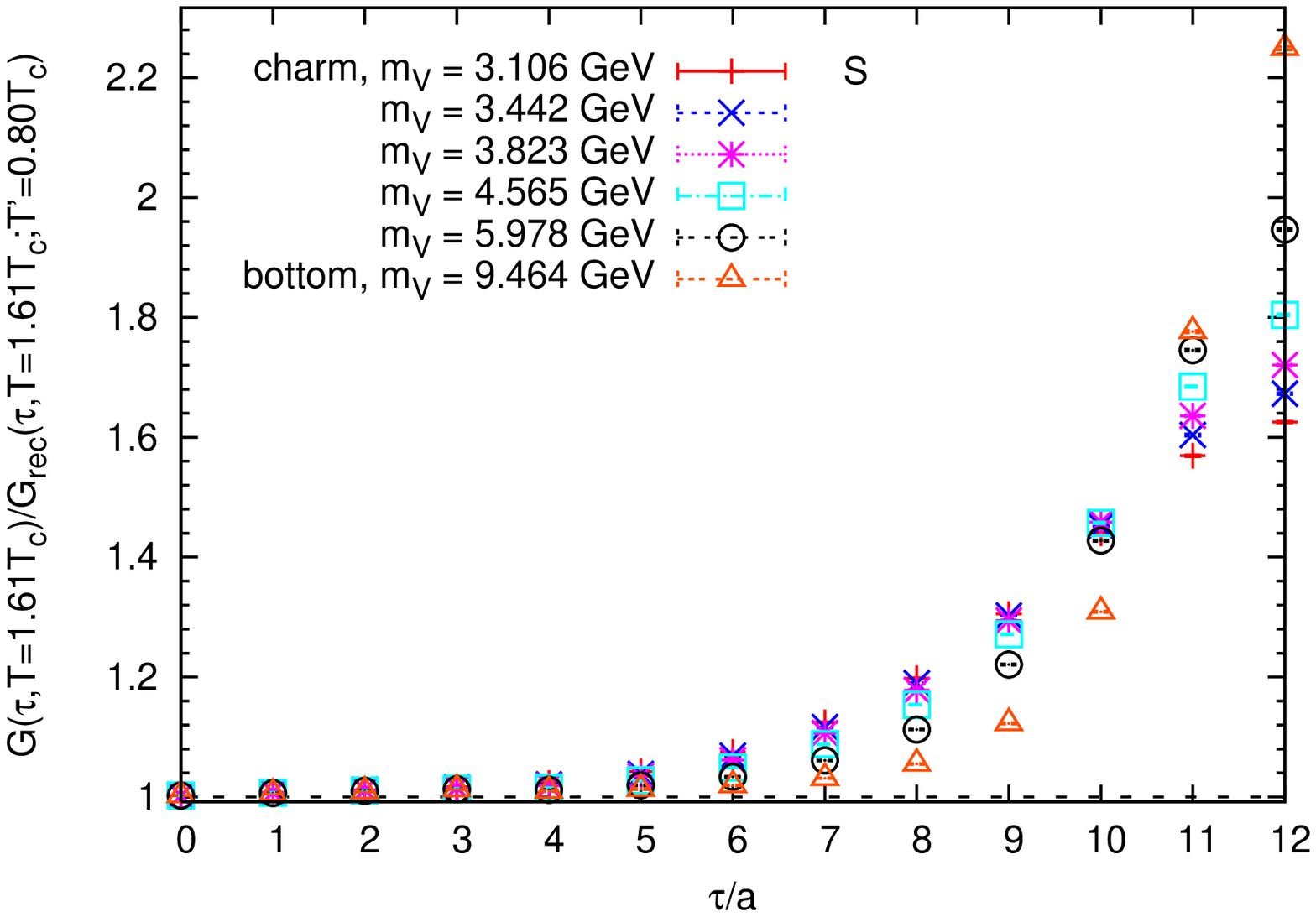}
  \includegraphics[width=49mm, angle=-90]{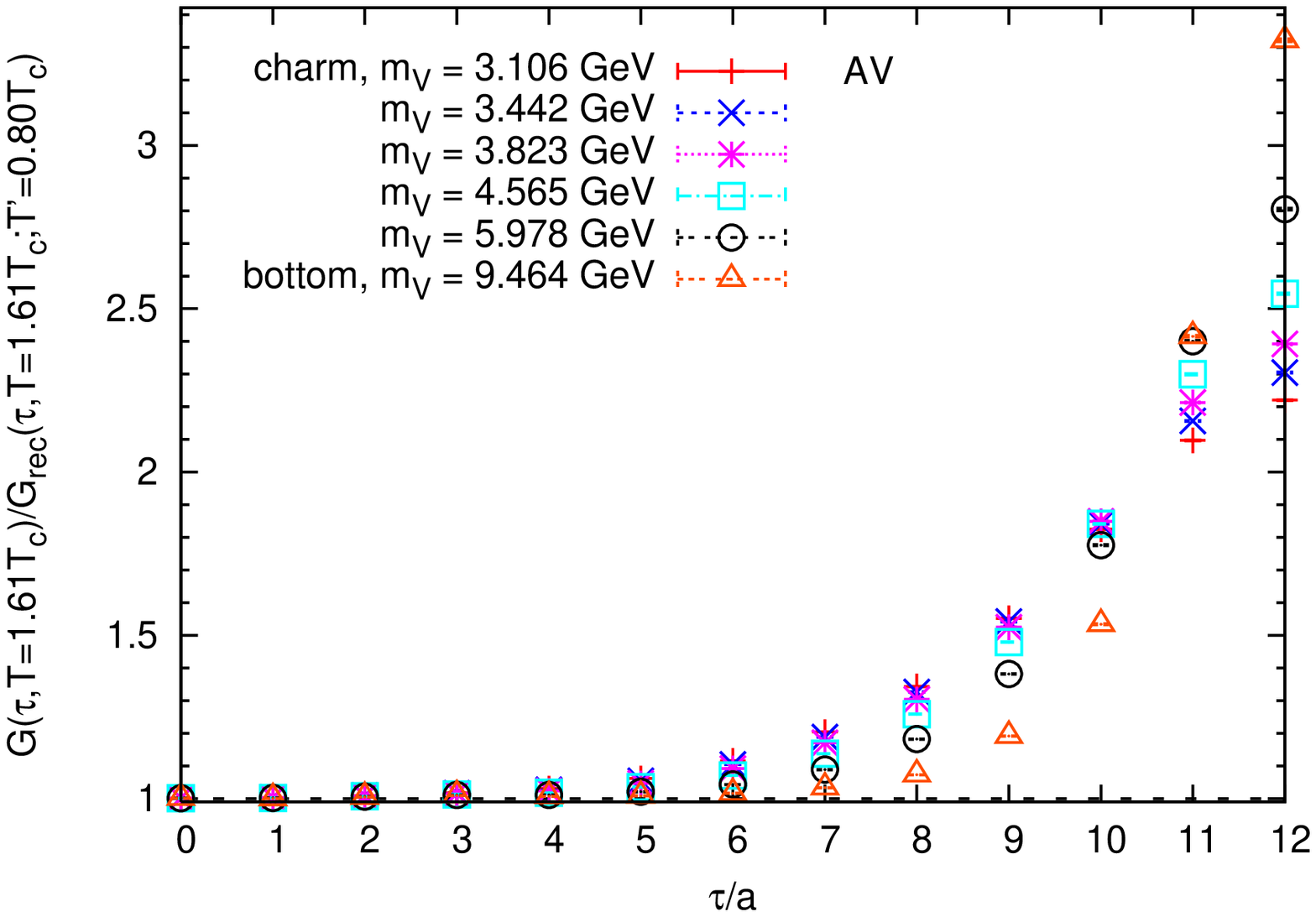}
  \caption{Quark mass dependence of the ratio of the temporal correlator at $T=1.61T_c$ to the corresponding reconstructed correlator given from
  	the $T=0.80T_c$ data for PS (top-left), V (top-right), S (bottom-left) and AV (bottom-right) channels. \label{rec_corr}}
  \vspace{-1em}
 \end{center}
\end{figure}

\begin{figure}[tbp]
 \begin{center}
  \includegraphics[width=49mm, angle=-90]{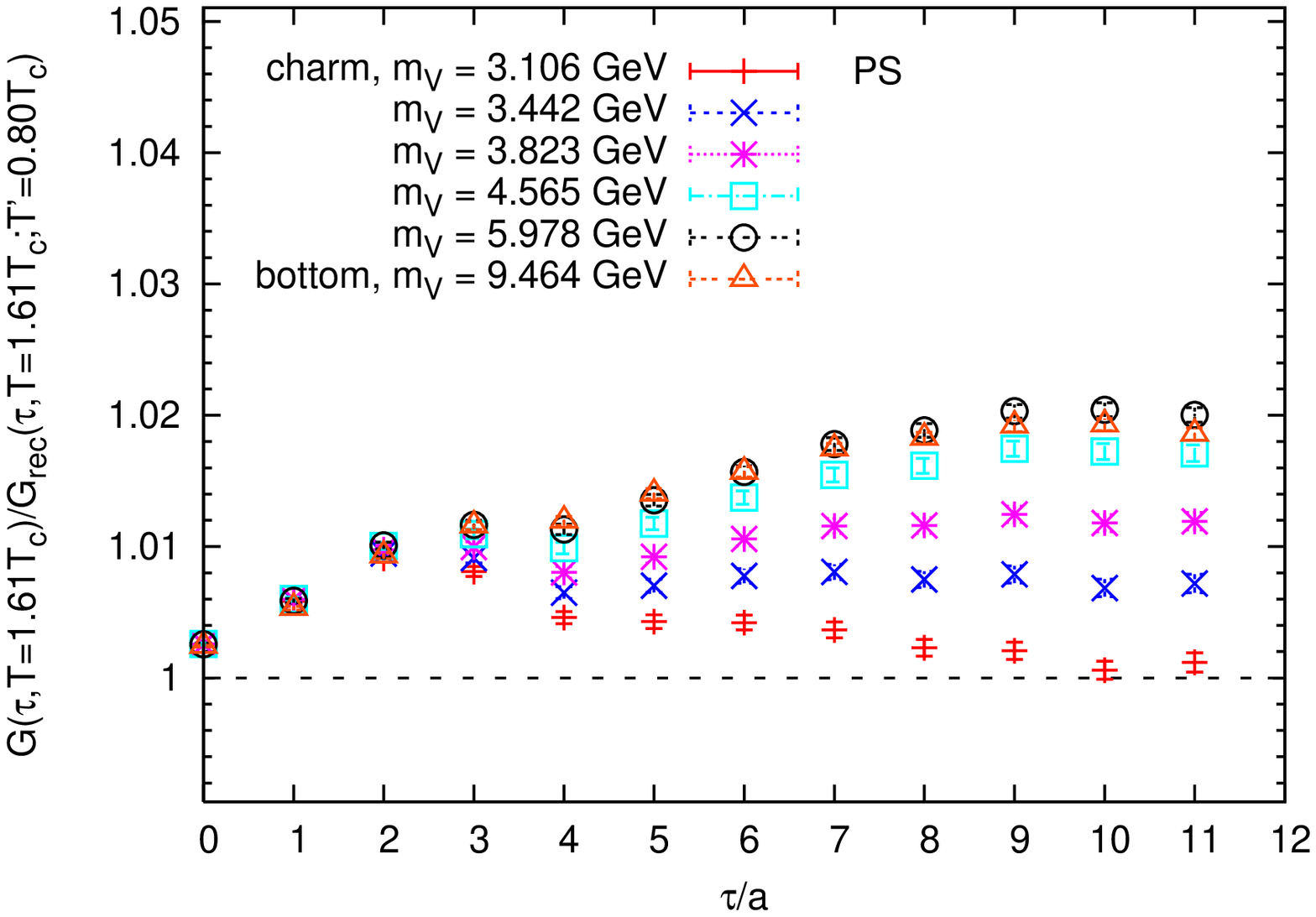}
  \includegraphics[width=49mm, angle=-90]{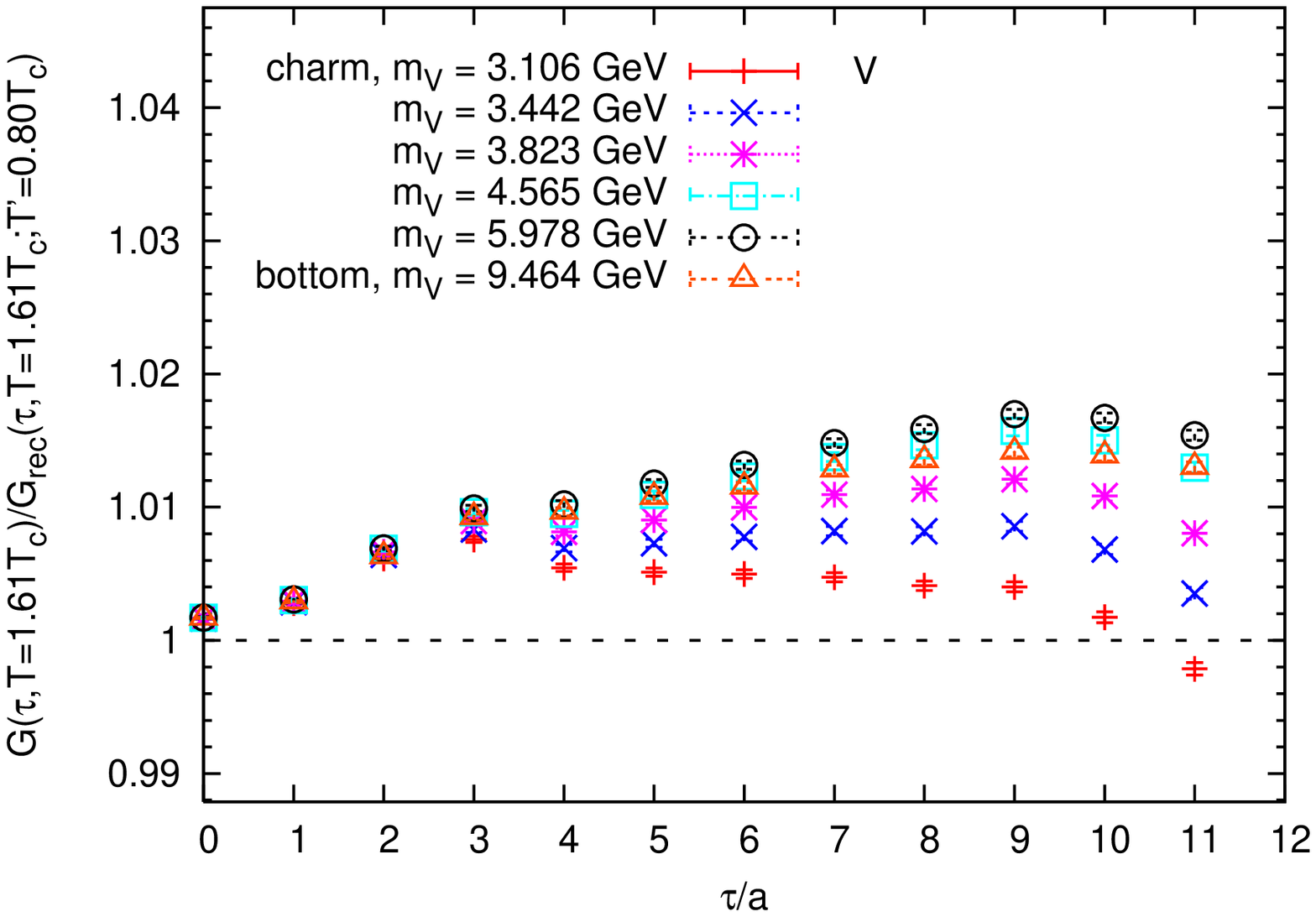}
  \includegraphics[width=49mm, angle=-90]{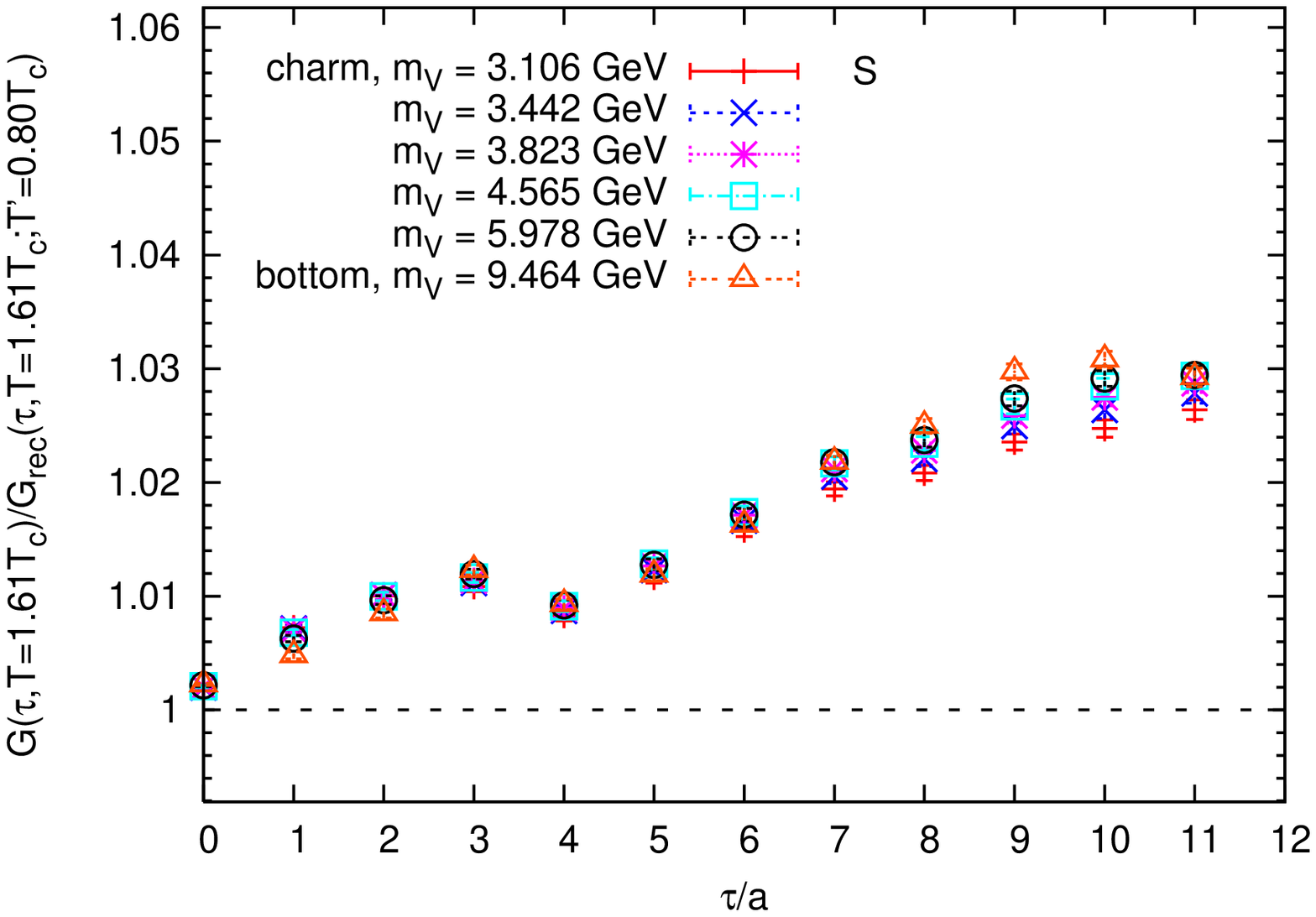}
  \includegraphics[width=49mm, angle=-90]{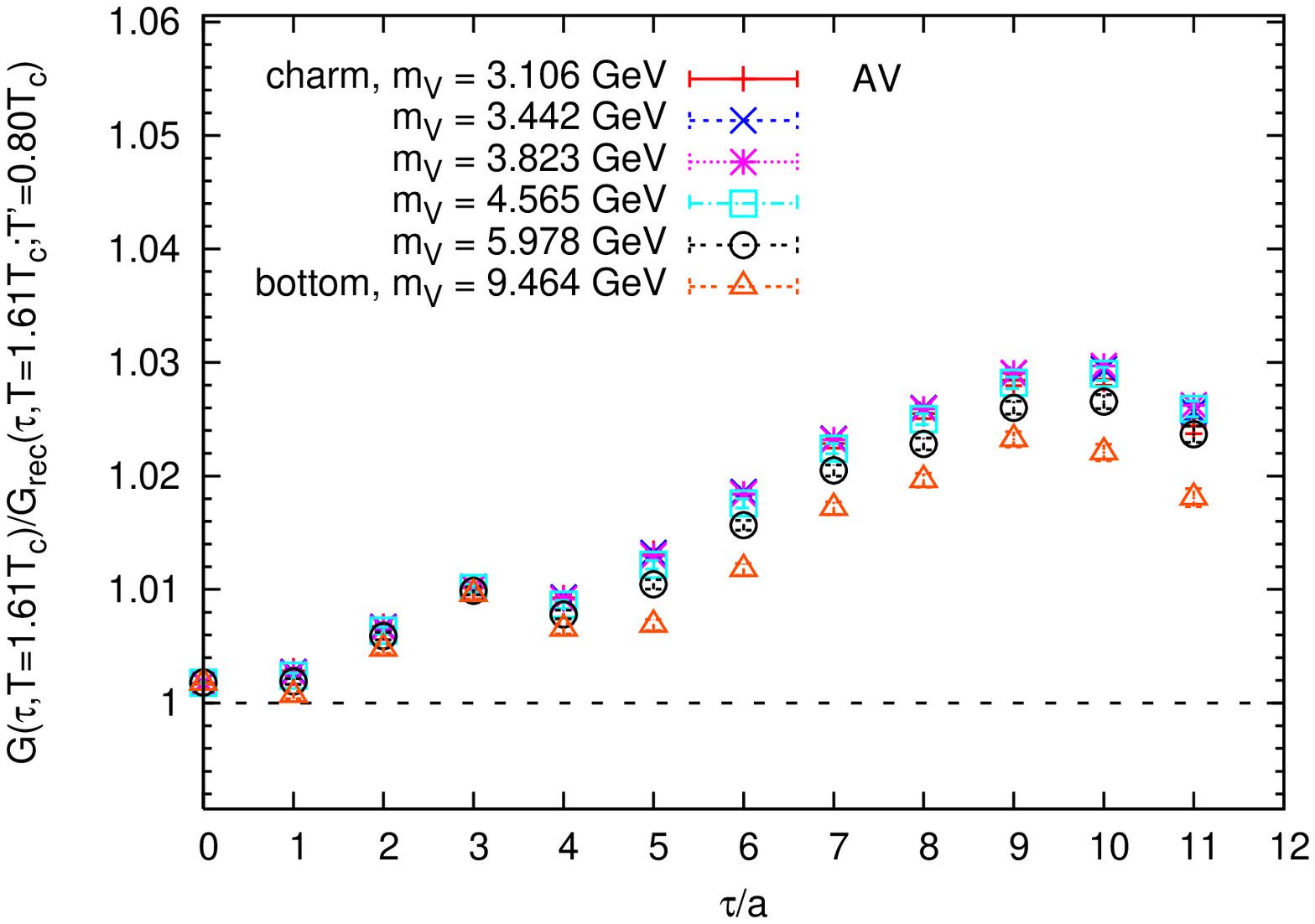}
  \caption{Same as Figure 2 but the midpoint subtraction technique is applied. \label{rec_corr_msb}}
  \vspace{-1em}
 \end{center}
\end{figure}

Figure \ref{rec_corr} shows quark mass dependence of the ratio of the temporal correlator at $T=1.61T_c$ to the corresponding reconstructed correlator given
by using the $T=0.80T_c$ data. For all channels the ratio has some quark mass dependence only at larger $\tau/a$ part, where small $\omega$ part of the
spectral function would be dominate. Since the small $\omega$ part of the spectral function is expected to consist of some bound state and transport peaks if they exist,
this quark mass dependence should correspond to the quark mass dependence for the modification of the bound state and transport peaks due to the thermal
effect. In this sense the bound state or transport peaks of the V, S and AV spectral functions seem to be strongly modified above $T_c$. Here, to separate the
$\omega=0$ contribution, which would be most dominant part of the transport peak, from the other part in the spectral function, we adopted the midpoint subtraction
technique \cite{Umeda:2007hy} to both the denominator and the numerator of the ratio. Figure \ref{rec_corr_msb} shows similar result to Figure \ref{rec_corr} but
the midpoint subtraction technique is applied. In this case the strong modification mentioned above disappears. This means that the most part of the strong modification
at small $\omega$ part in the spectral function is due to the transport contribution and the PS channel doesn't have such contribution.
Moreover, the S-wave channels have similar quark mass dependence to each other and the modification of the spectral function is larger for larger quark mass.
On the other hand, the both of P-wave channels have small quark mass dependence and the modification is large for all quark masses.

\begin{figure}[tbp]
 \begin{center}
  \includegraphics[width=49mm, angle=-90]{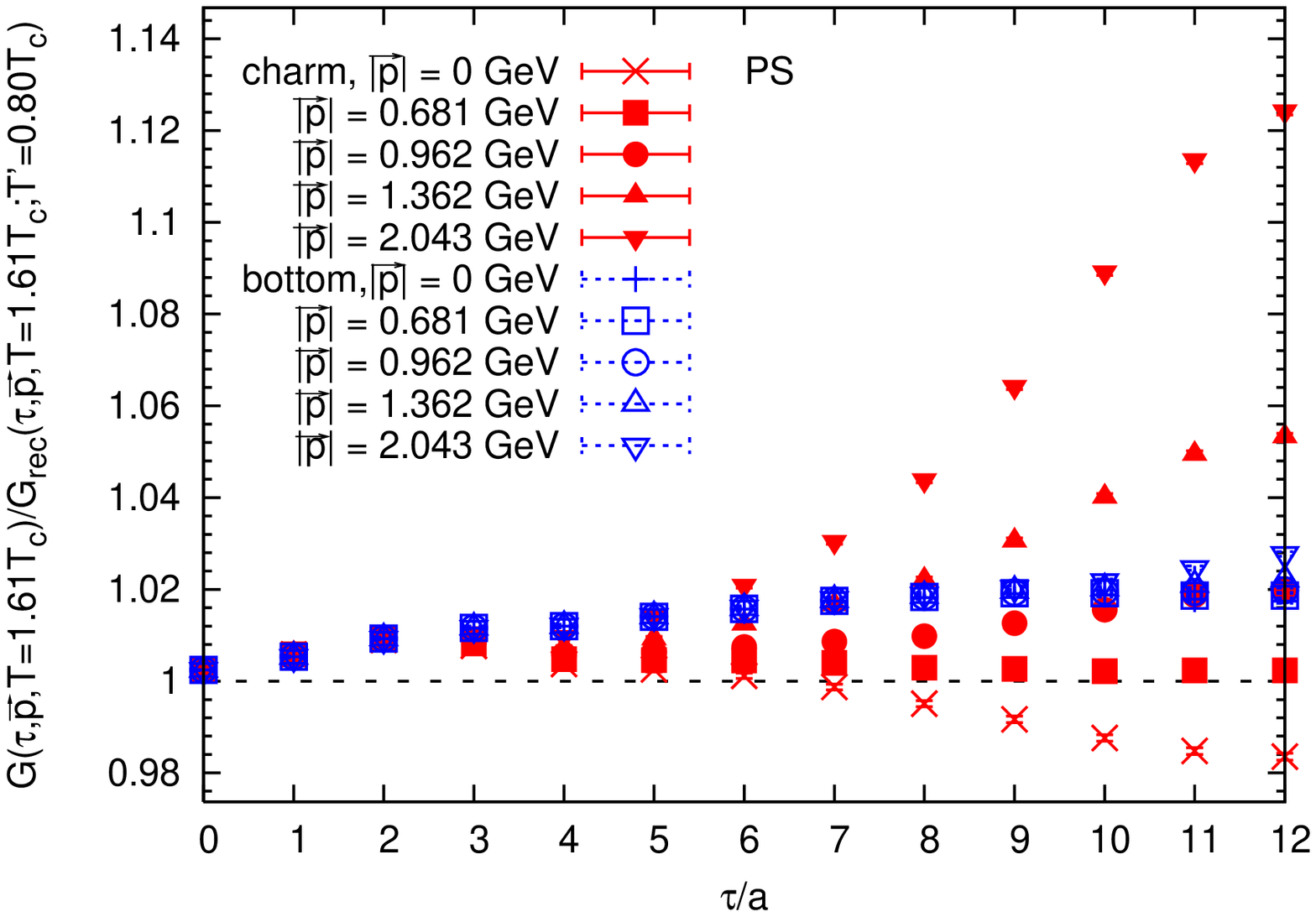}
  \includegraphics[width=49mm, angle=-90]{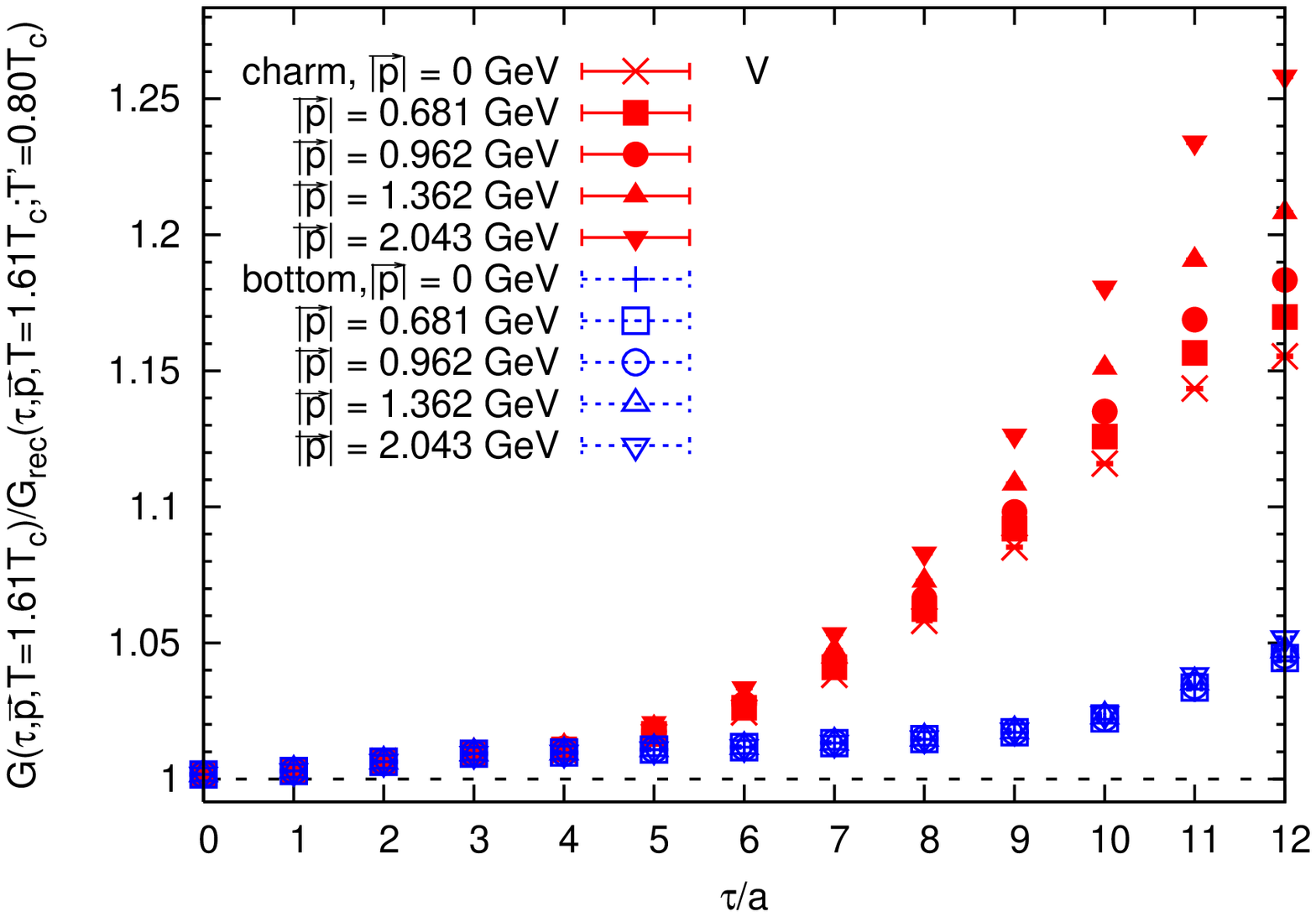}
  \includegraphics[width=49mm, angle=-90]{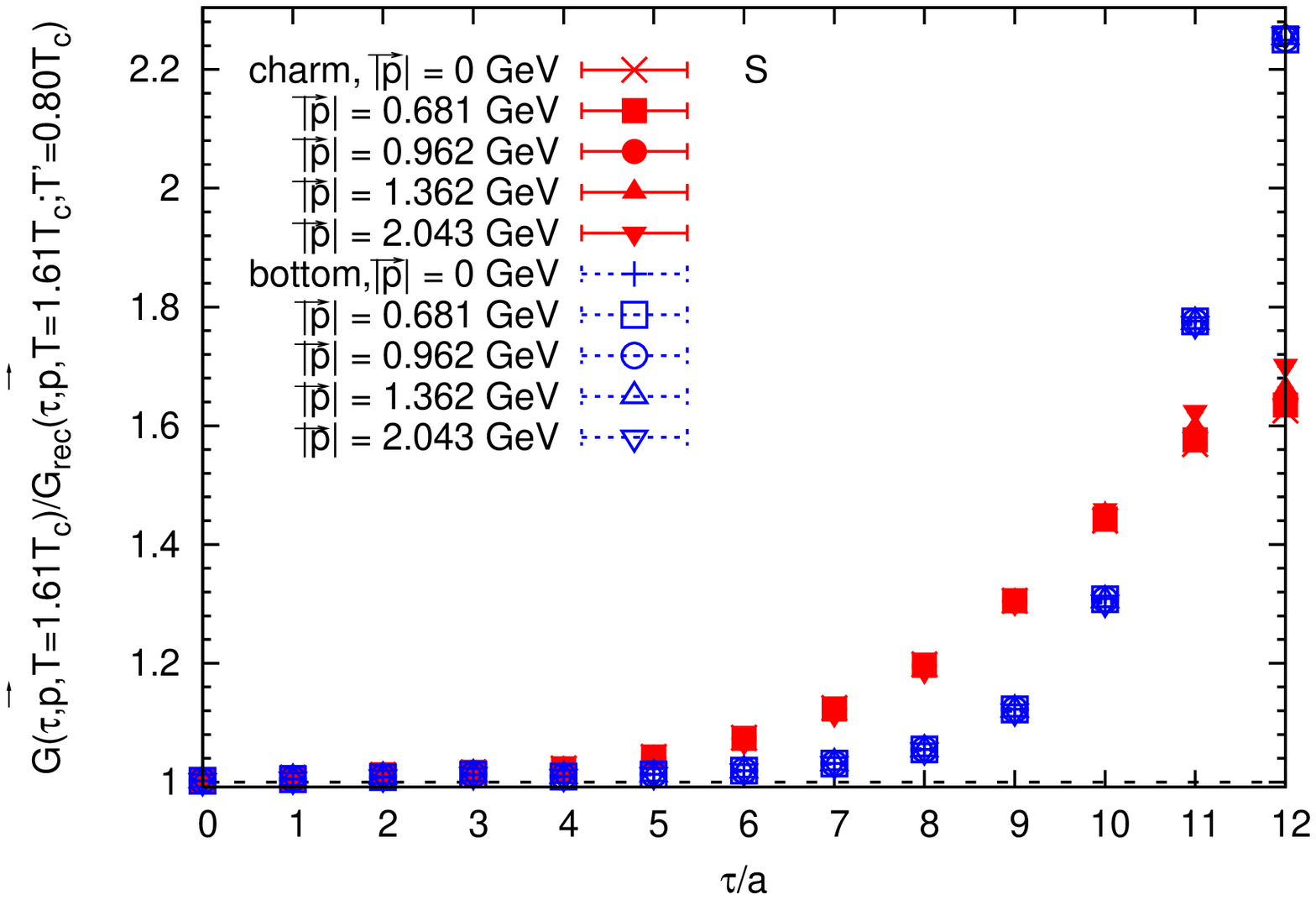}
  \includegraphics[width=49mm, angle=-90]{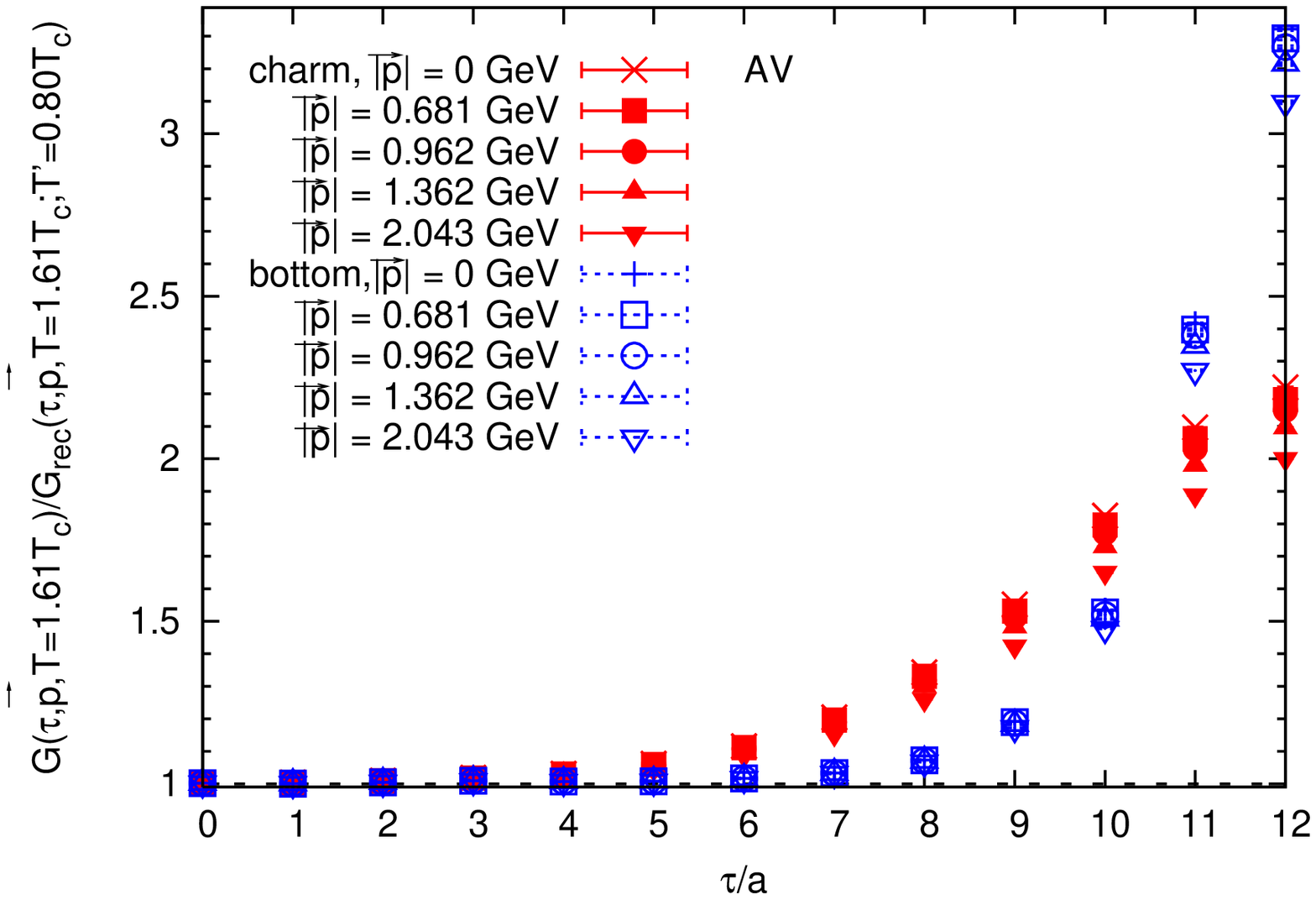}
  \caption{Momentum dependence of the same quantity shown in Figure 2. \label{rec_corr_MOMdep}}
  \vspace{-1em}
 \end{center}
\end{figure}

Finally, we consider non-zero momentum case, which is related to quarkonia moving in medium.
Momentum dependence of the same ratio discussed above is shown in Figure \ref{rec_corr_MOMdep}.
For the charmonia there is about 10--20\% momentum dependence at the largest $\tau/a$ for all channels
while the bottomonia have small momentum effect within the range of momentum investigated in this study,
except for the AV channel.

\section{Conclusions}
We studied quarkonium behavior at finite temperature in the region of the quark mass for charmonia to bottomonia
on large isotropic lattices. We investigated the screening mass and the reconstructed correlator.
The screening masses for the S-wave channels had larger thermal effect for lighter quark mass and it was negligibly small for
the bottomonia up to 1.21$T_c$. In case of the P-wave the screening mass was modified just above $T_c$
even for the bottomonia. According to the reconstructed correlator study at 1.61$T_c$, it was found that the spectral function
for V, S and AV channels had large transport contribution. The remaining part had larger thermal effect for larger quark mass
for the S-wave states but only small quark mass dependence for the P-wave states. The reconstructed correlator was also investigated at
finite momenta and the bottomonia except for the AV channel had quite small momentum dependence in the range of momentum we investigated.

Direct investigation of the spectral function, computing transport coefficients
and performing simulation on finer and larger lattices to take a continuum limit are our future plan. 

\acknowledgments{
I thank Olaf Kaczmarek for discussions.
This work has been supported in part by the European Union under grant 238353.
The numerical calculations have been performed on the Bielefeld GPU cluster and the OCuLUS Cluster
at The Paderborn Center for Parallel Computing in Germany.}


\begin{thebibliography}{99}
  \bibitem{Matsui:1986dk}
  T.~Matsui and H.~Satz,
  Phys.\ Lett.\ B {\bf 178}, 416 (1986).
  \bibitem{Arnaldi:2009ph} 
  R.~Arnaldi [NA60 Collaboration],
  Nucl.\ Phys.\ A {\bf 830}, 345C (2009)
  [arXiv:0907.5004 [nucl-ex]].
  \bibitem{Adare:2008qa} 
  A.~Adare {\it et al.}  [PHENIX Collaboration],
  Phys.\ Rev.\ Lett.\  {\bf 101}, 232301 (2008)
  [arXiv:0801.4020 [nucl-ex]].
  \bibitem{Abelev:2012rv} 
  B.~Abelev {\it et al.}  [ALICE Collaboration],
  Phys.\ Rev.\ Lett.\  {\bf 109}, 072301 (2012)
  [arXiv:1202.1383 [hep-ex]].
  \bibitem{Aad:2010aa} 
  G.~Aad {\it et al.}  [ATLAS Collaboration],
  Phys.\ Lett.\ B {\bf 697}, 294 (2011)
  [arXiv:1012.5419 [hep-ex]].
  \bibitem{Chatrchyan:2012np} 
  S.~Chatrchyan {\it et al.}  [CMS Collaboration],
  JHEP {\bf 1205}, 063 (2012)
  [arXiv:1201.5069 [nucl-ex]].
  \bibitem{Chatrchyan:2012lxa} 
  S.~Chatrchyan {\it et al.}  [CMS Collaboration],
  Phys.\ Rev.\ Lett.\  {\bf 109}, 222301 (2012)
  [arXiv:1208.2826 [nucl-ex]].
  \bibitem{Jakovac:2006sf} 
  A.~Jakovac, P.~Petreczky, K.~Petrov and A.~Velytsky,
  Phys.\ Rev.\ D {\bf 75}, 014506 (2007)
  [hep-lat/0611017].
  \bibitem{Aarts:2007pk} 
  G.~Aarts, C.~Allton, M.~B.~Oktay, M.~Peardon and J.~-I.~Skullerud,
  Phys.\ Rev.\ D {\bf 76}, 094513 (2007)
  [arXiv:0705.2198 [hep-lat]].
  \bibitem{Ohno:2011zc} 
  H.~Ohno {\it et al.}  [WHOT-QCD Collaboration],
  Phys.\ Rev.\ D {\bf 84}, 094504 (2011)
  [arXiv:1104.3384 [hep-lat]].
  \bibitem{Ding:2012sp} 
  H.~T.~Ding, A.~Francis, O.~Kaczmarek, F.~Karsch, H.~Satz and W.~Soeldner,
  Phys.\ Rev.\ D {\bf 86}, 014509 (2012)
  [arXiv:1204.4945 [hep-lat]].
  \bibitem{Aarts:2010ek} 
  G.~Aarts, S.~Kim, M.~P.~Lombardo, M.~B.~Oktay, S.~M.~Ryan, D.~K.~Sinclair and J.~-I.~Skullerud,
  Phys.\ Rev.\ Lett.\  {\bf 106}, 061602 (2011)
  [arXiv:1010.3725 [hep-lat]].
  \bibitem{Aarts:2012ka} 
  G.~Aarts, C.~Allton, S.~Kim, M.~P.~Lombardo, M.~B.~Oktay, S.~M.~Ryan, D.~K.~Sinclair and J.~-I.~Skullerud,
  JHEP {\bf 1303}, 084 (2013)
  [arXiv:1210.2903 [hep-lat]].
  \bibitem{Beringer:1900zz} 
  J.~Beringer {\it et al.}  [Particle Data Group Collaboration],
  Phys.\ Rev.\ D {\bf 86}, 010001 (2012).
  \bibitem{Umeda:2007hy} 
  T.~Umeda,
  Phys.\ Rev.\ D {\bf 75}, 094502 (2007)
  [hep-lat/0701005].
\end{thebibliography}
\end{document}